\documentclass[11pt]{article}
\usepackage[symbol]{footmisc}

\usepackage[left=2cm,top=2.5cm,right=2cm,bottom=2.5cm]{geometry}
\usepackage{amsmath, amssymb}

\usepackage{graphicx}
\usepackage{graphics}
\usepackage{epstopdf}
\usepackage{subfigure}
\usepackage{amsfonts}
\usepackage{sectsty}
\usepackage{sectsty}
\usepackage{hyperref}
\usepackage{cite}
\usepackage{lineno}

\begin{document}
	
	\begin{center}
	\large{\bf{FRW Cosmology in $f(Q,T)$ Gravity}} \\

	\vspace{5mm}
	\normalsize{Nisha Godani$^1$ and Gauranga C. Samanta$^{2}{}$ }\\
	\normalsize{$^1$Department of Mathematics, Institute of Applied Sciences and Humanities\\ GLA University, Mathura, Uttar Pradesh, India\\
		$^2$P. G. Department of Mathematics, Fakir Mohan University, Balasore, Odisha, India}\\
	\normalsize {nishagodani.dei@gmail.com\\gauranga81@gmail.com}
\end{center}

\begin{abstract}
\noindent
In this paper, we considered the study of Friedmann-Robertson-Walker (FRW) model in the framework of $f(Q,T)$ gravity, an extension of symmetric teleparallel gravity, recently defined by Y. Xu et al. \cite{Xu}. The non-linear model $f(Q,T)=-\alpha Q-\beta T^2$, where $\alpha>0$ and $\beta>0$ are constants, is taken into account. The equation of state of perfect fluid is assumed and 31 points of Hubble data are used to constrain the value of model parameter. To explore the evolution of the universe, the numerical solutions of cosmological implications such as Hubble parameter, deceleration parameter, apparent magnitude and luminosity distance are determined and the energy conditions are examined. The theoretical results of Hubble parameter are compared with $\Lambda$CDM model. Further, 57 Supernova data (42 from Supernova cosmology project and 15 from Cal\'{a}n/ Tolono supernova survey) are also used to have consistent results of apparent magnitude and luminosity distance.
\end{abstract}

\textbf{Keywords:}  $f(Q,T)$ gravity; Hubble parameter; Apparent magnitude; Luminosity distance.
\section{Introduction}
The recent observational progresses in cosmology have provided that our universe already entered into an accelerated expansion stage, and some exotic form of matter present in the universe could be the cause of this expansion\cite{Riess09, Perlmutter65, Ade2016, Akrami2018, Aghanim2018}. Subsequently, the same observations specify that around $95-96\%$ of the matter content of the universe is in the form of two types of mysterious components, called dark matter and dark energy, respectively, and nearly $4-5\%$ is in the form of baryonic matter\cite{Peebles2003, Padmanabhan2003}.

After several achievements and remarkable success of standard general relativity still, it may not be adequate completely to elucidate gravitational phenomena on galactic and cosmological ranges. So, general relativity may not be the final theory of the gravitational force, since it cannot provide satisfactory explanations to the fundamental problems in present day cosmology is challenged with: dark energy and dark matter problem respectively.
Therefore, the modification of general relativity is required to explore several aspects in modern cosmology.
There are many different approaches have been proposed at the classical level to explain the observational results of cosmology. Nevertheless, an adequate theory of gravity has yet to come. One of the simplest way of modifying Einstein's gravity is to introduce an arbitrary function of the Ricci scalar $R$ into the gravitational action \cite{Buchdahl1970}, which becomes $S=\frac{1}{2\kappa^2}\int f(R)\sqrt{-g}d^4x+\int L_m \sqrt{-g}d^4x $.
A second method to modify the Einstein gravity is to assume the existence of a non-minimal coupling between matter and geometry. This way of research provides us to different classes of gravitational theories, called $f(R, L_m)$ gravity \cite{Bertolami2007, Harko2008}, with action given by $S=\int f(R, L_m)\sqrt{-g}d^4x$, and $f(R, T)$ gravity \cite{Harko}, with action given by $S=\int f(R, T)\sqrt{-g}d^4x$, where
$T$ is the trace of the energy-momentum tensor, respectively.
For wide-range of studies and discussions on modified gravity theories and of their implications see \cite{Nojiri2003prd, Nojiri2004grg, Nojiri2006prd05, Nojiri2007ijgmmp, Nojiri2007jpa, Nojiri2008ptps, Nojiri2008prd02, Nojiri2008plb, Bambajcap, Lobo2009sig, Bamba2010epjc, Bamba2010mpla, Elizalde2010epjc, Capozziello2011pr, Olmo2011ijmpd, Bamba2011plb51, Nojiri2011pr, Bamba2012ass, Bambaprd012, Bamba2012prd036, Elizalde2012epjc43, samanta2013ijtp1, Samanta2013ijtp, Bamba2014plb, Bamba2015epjc, Bamba2015s, Astashenok2015, Yousaf2016prd, Nojiri2017pr, Godani2019ijmpd, Mishra21018mpla, Yousaf2019pdu, Shah2019epjc, Samanta2019epjc, godani2019cjp, Shah2020ijmpa, Tripathy2020, Godani2020cjp, Elizalde2020pdu, Godani2020epjc1, Bhatti2020pdu}.
Nevertheless, Harko et al \cite{Harko2018epjc1} extended symmetric teleparallel gravity by introducing nonmetricity $Q$ is nonminimally coupled to matter Lagrangian, this leads to the nonconservation of the energy-momentum tensor, and subsequently the appearance of an extra force in the geodesic equation of motion. A Lagrangian of the form $L=f_1(Q)+f_2(Q)L_m$ was considered, where $f_1$ and $f_2$ are generic functions of $Q$, and $L_m$
is the matter Lagrangian.

Recently, Y. Xu et al. \cite{Xu} investigated another extension of $f(Q)$ gravity, which is based on the nonminimal coupling between the nonmetricity $Q$ and the trace $T$ of the matter energy-momentum tensor. In precisely, the Lagrangian density of the gravitational field is defined by a general function of both $Q$ and $T$, so that $L=f(Q, T)$. Subsequently, Xu et al \cite{Xu2020epjc} studied $f(Q, T)$ gravity by taking simple functional forms of the function
$f(Q, T)$, and compare their results with the standard $\wedge$CDM model. Arora et al \cite{Arora2020pdu} investigated $f(Q, T)$ gravity with observational constraints by considering the simple form of $f(Q, T)$, i. e. $f(Q, T)=mQ^n+bT$, and specific form of scale factor.

In the light of the above discussion, the main objective of this paper is to study the viability of $f(Q, T)$ model with observational data by taking non-linear form of $f(Q, T)$, i. e. $f(Q, T)=-\alpha-\beta T^2$. Eventually, we compare our results with $\wedge$CDM model.

\section{$f(Q, T)$ gravity and field equations}

Recently, the $f(Q,T)$ theory of gravity has been introduced by Y Xu et al. \cite{Xu}. The gravitational action in $f(Q,T)$ gravity is defined as
\begin{equation}\label{action2}
S=\dfrac{1}{16\pi}\int f(Q,T)\sqrt{-g}d^4x +\int \sqrt{-g}\mathcal{L} d^4x,
\end{equation}
where $Q$ is the non-metricity, $T$ is the trace of stress energy tensor, $f(Q,T)$ is a function of $Q$ and $T$, $L$ is the matter Lagrangian and $g=det(g_{\mu\nu})$. The non-metricity $Q$ is given by
\begin{equation}
Q\equiv-g^{\mu\nu}(L^\alpha_{\beta\mu}L^\beta_{\nu\alpha}-L^\alpha_{\beta\alpha}L^\beta_{\mu\nu}),
\end{equation}
where $L^\alpha_{\beta\gamma}$ denotes the deformation tensor
\begin{equation}
L^{\alpha}_{\beta\gamma}=-\frac{1}{2}g^{\alpha\lambda}(\bigtriangledown_\gamma g_{\beta\lambda}+ \bigtriangledown_\beta g_{\gamma\lambda} - \bigtriangledown_\lambda g_{\beta\gamma}).
\end{equation}
The trace of stress energy tensor is defined as
\begin{equation}\label{}
T_{\mu\nu}=-\frac{2\delta (\sqrt{-g}\mathcal{L})}{\sqrt{-g}\delta g^{\mu\nu}}.
\end{equation}
Then the field equations by varying the gravitational action are obtained as
\begin{equation}\label{fe}
8\pi T_{\mu\nu}=-\frac{2}{\sqrt{-g}}\bigtriangledown_\alpha(f_Q\sqrt{-g}P^{\alpha}_{\mu\nu}-\frac{fg_{\mu\nu}}{2}+f_T(T_{\mu\nu}+\Theta_{\mu\nu})-f_Q(P_{\mu\alpha\beta}Q_{\nu}^{\alpha\beta}-2Q_{\mu}^{\alpha\beta}P_{\alpha\beta\nu})),
\end{equation}
where $\Theta_{\mu\nu}\equiv g^{\alpha\beta}\frac{\delta T_{\alpha\beta}}{\delta g^{\mu\nu}}$ and $P^{\alpha}_{\mu\nu}$ stands for the superpotential of the model \cite{Xu}. \\

Now, we consider flat, isotropic and homogeneous FRW metric
\begin{equation}\label{metric}
ds^2=-dt^2+a^2(t)(dx^2+dy^2+dz^2),
\end{equation}
where $a(t)$ denotes the scale factor. The expansion rate is defined in terms of scale factor as $H\equiv\frac{\dot{a}}{a}$. The function $H$ is called Hubble parameter and non-metricity $Q=6H^2$ \cite{Xu}.

From \eqref{fe} and \eqref{metric}, the field equations are obtained as
\begin{equation}\label{fe1}
8\pi\rho=\frac{f}{2}-6FH^2-\frac{2\tilde{G}}{1+\tilde{G}}(\dot{F}H+F\dot{H})
\end{equation}
and
\begin{equation}\label{fe2}
8\pi p=-\frac{f}{2}+6FH^2+2(\dot{F}H+F\dot{H}).
\end{equation}
From Equations \eqref{fe1} and \eqref{fe2}, the evolution equation for $H$ comes out to be
\begin{equation}\label{evolution}
\dot{H} + \frac{\dot{F}H}{F} = \frac{4\pi}{F}(1+\tilde{G})(\rho+p).
\end{equation}

Introducing the effective pressure and effective energy density, the field equations reduce to

\begin{equation}
3H^2=8\pi \rho_{eff}=\frac{f}{4F}-\frac{4\pi}{F}\Big[(1+\tilde{G})\rho+\tilde{G} p\Big],
\end{equation}

\begin{equation}
2\dot{H}+3H^2=-8\pi p_{eff}=\frac{f}{4F}- \frac{2\dot{F}H}{F}+\frac{4\pi}{F}\Big[(1+\tilde{G})\rho+(2+\tilde{G}) p\Big].
\end{equation}

The effective thermodynamic quantities satisfy conservation equation  which are as followed: \begin{equation}
\dot{\rho_{eff}} +3H(\rho_{eff}+p_{eff})=0.
\end{equation}

\section{Estimation of Cosmological Parameters and Energy Conditions}
In this section,  the numerical solutions for cosmological parameters such as Hubble parameter and deceleration parameter are determined in the context of $f(Q,T)$ gravity with non-linear form of $f(Q,T)$ function as $f(Q,T)=-\alpha Q-\beta T^2$, where $\alpha>0$ and $\beta>0$ are constants \cite{Xu}. Further, the nature of energy conditions which include null energy condition (NEC), weak energy condition (WEC), strong energy condition (SEC) and dominant energy condition (DEC) is examined.

For the function $f(Q,T)=-\alpha Q-\beta T^2$,  $F=\frac{\partial f}{\partial Q}=-\alpha$, $8\pi\tilde{G}=-2\beta T=2\beta\rho(t)$. We suppose the equation of state of perfect fluid given by $p=\omega \rho$, where $p$, $\rho$ and  $\omega$  denote pressure, density and equation of state parameter respectively.

\subsection{Energy Density}
From Equations \eqref{fe1} and \eqref{evolution}, the energy density is
\begin{eqnarray}
\rho(t)&=&\frac{f-12H^2(t)F}{16\pi((1+\omega)\tilde{G}+1)}\nonumber\\
    &=&\frac{6\alpha H^2(t)-\beta \rho^2(-1+3\omega)^2}{16\pi-4\beta\rho(t)(1+\omega)(-1+3\omega)}.
\end{eqnarray}
This gives the physical solution as
\begin{eqnarray}\label{r}
\rho(t)&=&\frac{8\pi\Big(-1\pm \sqrt{1+\frac{3}{32\pi^2}\alpha\beta H^2(t)(1-3\omega)(5+\omega)}\Big)}{\beta(1-3\omega)(5+\omega)}.
\end{eqnarray}
If $\frac{3}{32\pi^2}\alpha H^2(t)\beta(1-3\omega)(5+\omega)<<1$, then the expansion of  square root in \eqref{r} gives $\rho(t)\propto H^2(t)$.

\subsection{Hubble Parameter}
Further, the evolution equation for the Hubble function comes out to be
\begin{eqnarray}\label{H(t)}
\frac{dH}{dt}&=&\frac{(1+\omega)}{-\alpha}\Bigg[4\pi+\beta(1-3\omega)\frac{8\pi\Big(-1\pm \sqrt{1+\frac{3}{32\pi^2}\alpha H^2(t)\beta(1-3\omega)(5+\omega)}\Big)}{\beta(1-3\omega)(5+\omega)}
\Bigg]\nonumber\\
&\times&\frac{8\pi\Big(-1\pm \sqrt{1+\frac{3}{32\pi^2}\alpha H^2(t)\beta(1-3\omega)(5+\omega)}\Big)}{\beta(1-3\omega)(5+\omega)}.
\end{eqnarray}
Rescaling the Hubble function as
\begin{eqnarray}\label{scale}
H(t)=\sqrt{\frac{32\pi^2}{3\alpha\beta(1-3\omega)(5+\omega)}}h(t),
\end{eqnarray}
the equation \eqref{H(t)} takes the form of differential equation
\begin{eqnarray}\label{h(t)}
\frac{dh}{dt}&=&-k(-1+\sqrt{1+h^2(t)})\Big(1+\frac{2\sqrt{1+h^2(t)}}{5+\omega}\Big),
\end{eqnarray}
where $k=\sqrt{\frac{96\pi^2(1+\omega)^2}{\alpha\beta(1-3\omega)(5+\omega)}}$.
Converting the equation \eqref{h(t)} in terms of redshift, we have
\begin{eqnarray}\label{h(z)}
(1+z)h(z)\frac{dh(z)}{dz}=k(-1+\sqrt{1+h^2(z)})\Big(1+\frac{2\sqrt{1+h^2(z)}}{5+\omega}\Big).
\end{eqnarray}
Let $v(z)=h^2(z)$. Then the equation \eqref{h(z)} reduces to
\begin{eqnarray}\label{u(z)}
(1+z)\frac{dv(z)}{dz}=k(-1+\sqrt{1+v(z)})\Big(1+\frac{2\sqrt{1+v(z)}}{5+\omega}\Big).
\end{eqnarray}
\noindent
Equation \eqref{u(z)} is dependent on parameters $k$ and $\omega$, and variable $z$. From \eqref{scale}, $\omega$ can have values between -5 and 1/3. Now, the question is: what should the value of model parameter $k$  be chosen? To determine the best suitable value of $k$, we have used 31 points of $H(z)$ values mentioned at the end of the article and obtained the value of the model parameter $k$ by minimizing the chi-square value by
\begin{eqnarray}\label{chi}
\chi^2_{OHD}(p_s)=\sum_{n=1}^{28}\frac{[H_{th}(p_s,z_n)-H_{ob}(z_n)]^2}{\sigma^2},
\end{eqnarray}
where $H_{th}$ and $H_{ob}$ stand for the theoretical and observational values of $H(z)$ respectively, $\sigma$ denotes the standard error in $H_{ob}$  and $p_s$ denotes the parameter space to be constrained. Using \eqref{chi}, the value of $k$ is obtained as 1.8. Then the numerical solution of $h(z)$ is presented in Fig. (1) by taking $\omega=-2, -1/3, 0.1, 0.25$ and $k=1.8$. For each value of $\omega$, $h(z)$ is found to be an increasing function of $z$. This implies that our universe is expanding. In Fig. (1), $h(z)$ corresponding to $\Lambda$CDM is also presented. It is observed that the values of $h(z)$ for $\omega<-1$ deviate from $\Lambda$CDM more in comparison of values corresponding to $\omega>-1$ which can also be seen in Fig. (1).

\subsection{Deceleration Parameter}
%

The deceleration parameter $q$ is a cosmological quantity which describes the accelerating or decelerating nature of the universe evolution. It is defined by
\begin{eqnarray}
q&=&-\frac{\dot{H}}{H^2}-1\nonumber\\
&=&(1+z)\frac{1}{H(z)}\frac{dH(z)}{dz}-1
\end{eqnarray}
By examining the energy conditions above, we obtained that $\omega \in (-1,1/3)$ for having the universe filled with ordinary matter. Then we calculated the numerical solution of $q(z)$ and checked its nature for this range of $\omega$. It is found that $-1<q(z)<0$ for $z\geq 0$ with $\omega \in (-1,1/3)$. It is also shown in Fig. (6) for $\omega=-1/3, 0.1$ and $0.25$. Thus, according to our model, the evolution of universe is in accelerating   phase at present and it has been started many years ago.

\subsection{Luminosity Distance}	

\noindent
The observations of type Ia Supernova \cite{Riess09, Perlmutter65} have declared the evolution of  the universe in an accelerating way.   The luminosity of an stellar objects defines the luminosity distance which has a significant role in studying the evolution of the universe. It is defined in terms of redshift by
\begin{eqnarray}\label{d1}
D_L&=&a_0c(1+z)\int_{t}^{t_0}\dfrac{dt}{a(t)}\nonumber\\
&=&c(1+z)\int_{0}^{z}\dfrac{dz}{H(z)},
\end{eqnarray}
where $c$ and $a_0$ are the speed of light and the present value of the scale factor respectively.
We have determined luminosity distance theoretically with respect to $z$ from \eqref{d1} and compared  with the corresponding results from 57 Supernova data (42 from Supernova cosmology project and 15 from Cal\'{a}n/ Tolono supernova survey). It is observed that the theoretical and observational values of $D_L$ coincide for each $\omega\in(-1,1/3)$ and some values of $z$. It is plotted in Fig. (7) for $\omega=-1/3, 0.1$ and $0.25$. In this figure, the observational values of $D_L$ are also marked and found to be consistent with the values of $D_L$ obtained by solving \eqref{d1}.

\subsection{Apparent Magnitude}	
Further, the apparent magnitude of a light source is defined in terms of luminosity distance by the relation
\begin{equation}\label{mm}
m-M=5log_{10}\Big(\dfrac{D_L}{Mpc}\Big)+25,
\end{equation}
where  $m$ means apparent magnitude and $M$ means absolute magnitude.

For lower redshift, $D_L$ is
\begin{equation}\label{d}
D_L=\dfrac{cz}{H_0}.
\end{equation}
Using Eq.(\ref{d}) and substituting $z=0.026$ and $m=16.08$ in (\ref{mm}),
\begin{equation}\label{M}
M=5log_{10}\left(\dfrac{H_0}{0.026c}\right)-8.92.
\end{equation}
From Eqs. (\ref{mm}) \& (\ref{M}),
\begin{eqnarray}\label{m1}
m&=&16.08+5log_{10}\left(\dfrac{D_LH_0}{0.026c}\right).
\end{eqnarray}
By Equations \eqref{d1} and \eqref{m1},
\begin{eqnarray}\label{m}
	m &= &16.08+5log_{10}\left(\dfrac{(1+z)H_0}{0.026}\int_{0}^{z}\dfrac{dz}{H(z)}\right).
	\end{eqnarray}
Using Eq. \eqref{m}, we have obtained apparent magnitude with respect to redshift and compared  with the corresponding results. For each $\omega\in(-1,1/3)$, it  increases and coincides with the observational values of $m$ obtained from 57 Supernova data (42 from Supernova cosmology project and 15 from Cal\'{a}n/ Tolono supernova survey). In Fig. (8), the  theoretical values of $m$ are observed to be in good agreement with the observational values of $m$.

\begin{figure}[h!]\label{h}
	\begin{center}
		\includegraphics[scale=.5]{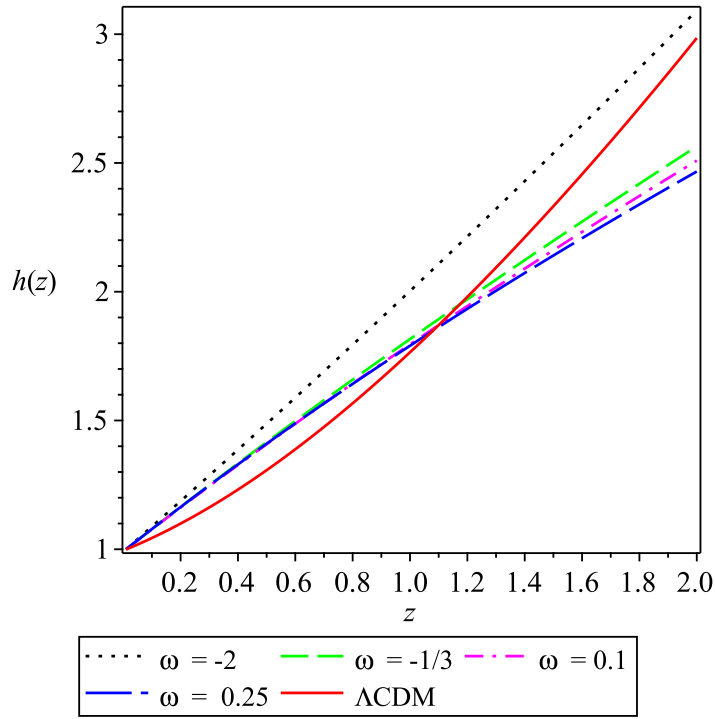}
		\caption{In this figure, the numerical solution of function $h(z)$ is plotted with respect to redshift $z$ for $\omega = -2, -1/3, 0.1$ and 0.25. It is also plotted for $\Lambda$CDM model which is closed to the values of $h(z)$ corresponding to different $\omega$ with lower redshifts.}
	\end{center}
\end{figure}
\begin{figure}[h!]\label{rho}
	\begin{center}
		\includegraphics[scale=.5]{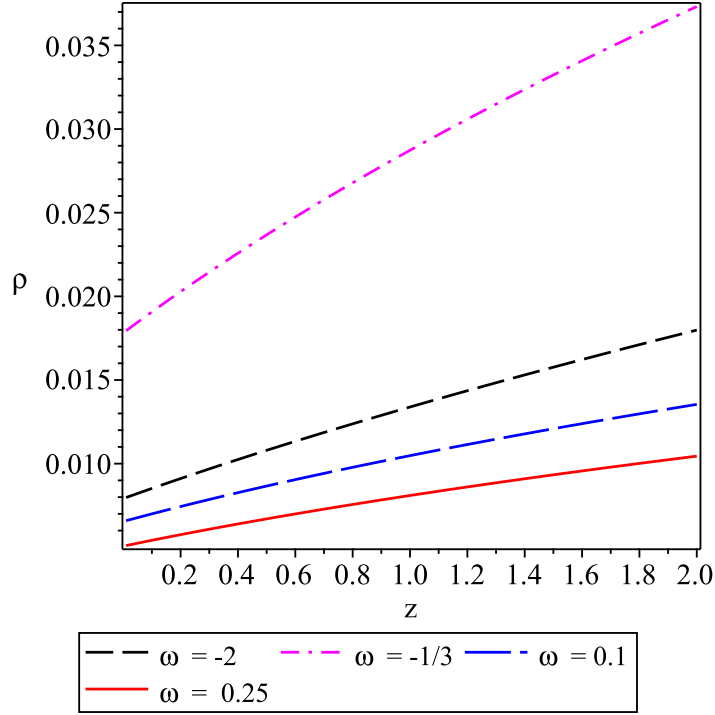}
		\caption{In this figure, the value of $\rho(z)$ is plotted with respect to redshift $z$ for $\omega = -2, -1/3, 0.1$ and 0.25. It is obtained to be positively increasing function for each $\omega$.}
	\end{center}
\end{figure}

\begin{figure}[h!]\label{nec}
	\begin{center}
		\includegraphics[scale=.5]{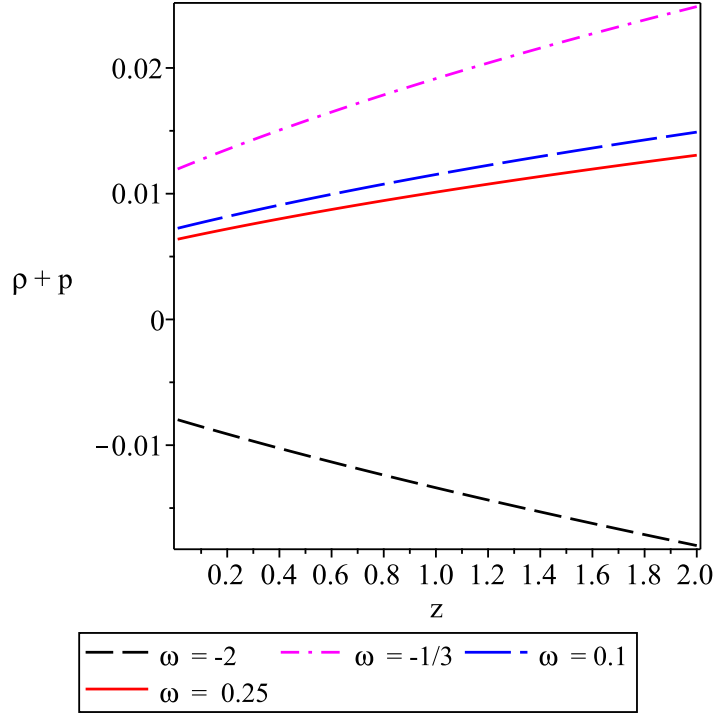}
		\caption{In this figure,  NEC term $\rho(z)+p(z)$ is plotted with respect to redshift $z$ for $\omega$ = -2, -1/3, 0.1 and 0.25. It is obtained to be positively increasing function   for $\omega$ =  -1/3, 0.1 and 0.25, and negatively decreasing function for $\omega$ = -2.}
	\end{center}
\end{figure}

\begin{figure}[h!]\label{sec}
	\begin{center}
		\includegraphics[scale=.5]{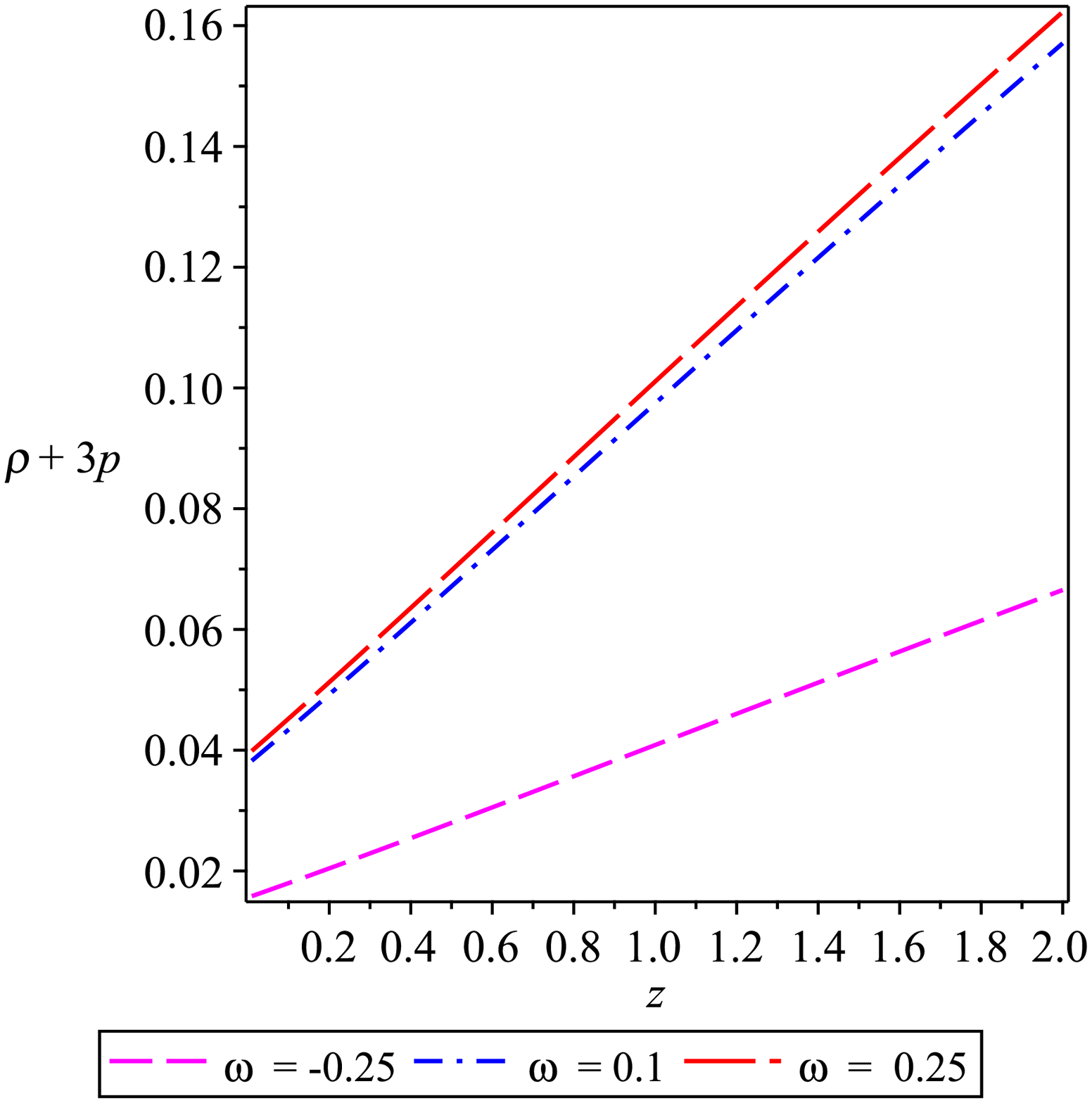}
		\caption{In this figure,  SEC term $\rho(z)+3p(z)$ is plotted with respect to redshift $z$ for $\omega$ = -0.25, 0.1 and 0.25. It is obtained to be positive and increasing function   for each $\omega$.}
	\end{center}
\end{figure}

\begin{figure}[h!]\label{dec}
	\begin{center}
		\includegraphics[scale=.5]{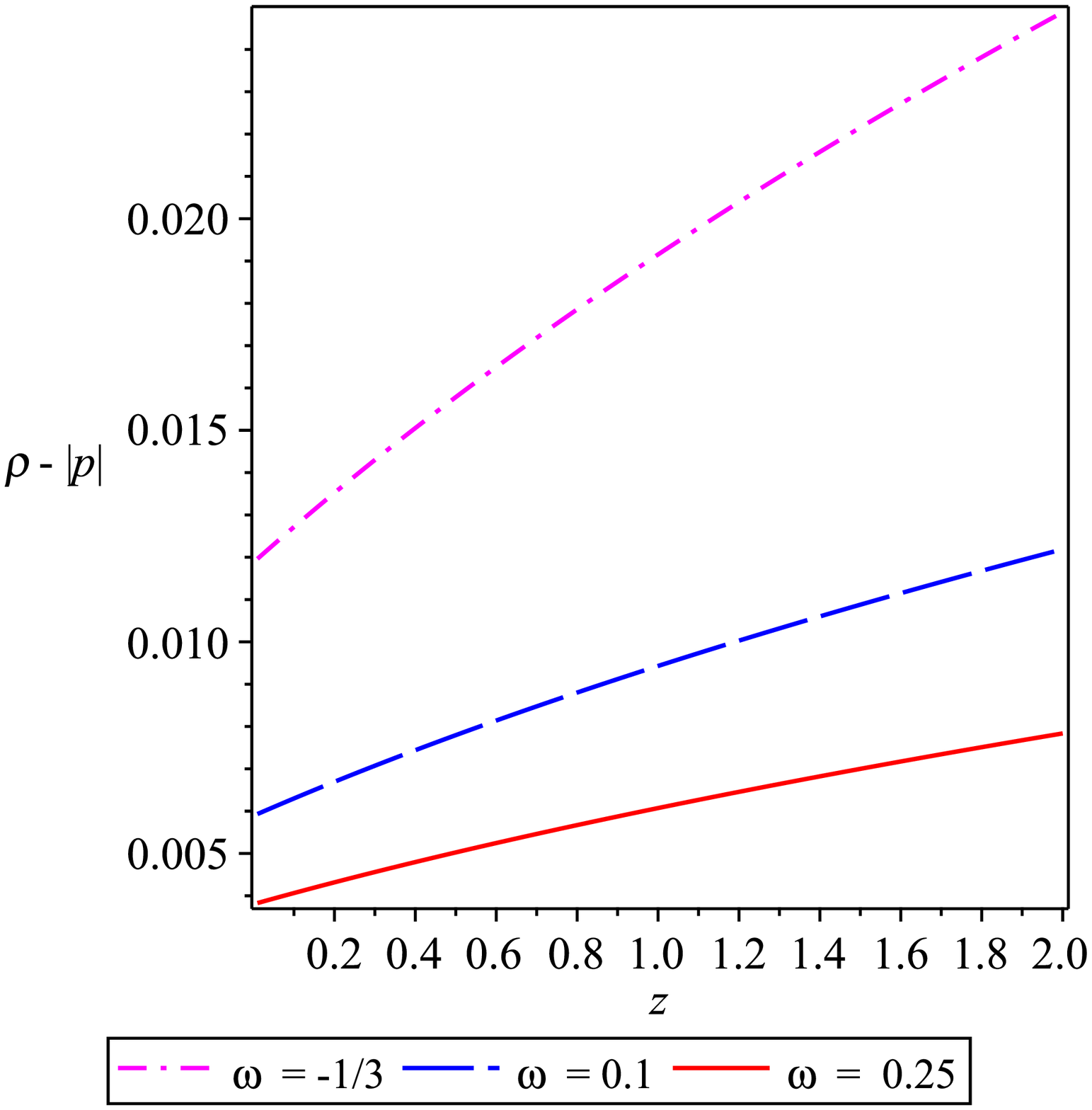}
		\caption{In this figure,  DEC term $\rho(z)-|p(z)|$ is plotted with respect to redshift $z$ for $\omega$ = -1/3, 0.1 and 0.25. It is obtained to be positive and increasing function   for each $\omega$.}
	\end{center}
\end{figure}

\begin{figure}[h!]\label{q}
	\begin{center}
		\includegraphics[scale=.5]{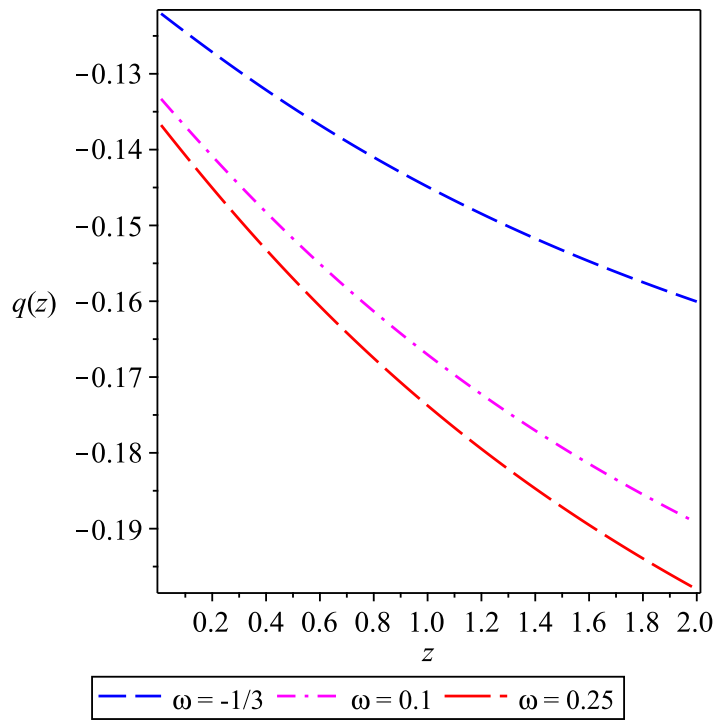}
		\caption{In this figure,  deceleration parameter  $q$ is plotted with respect to redshift $z$ for $\omega$ = -1/3, 0.1 and 0.25. It is obtained to be possess values between -1 and 0 for each $\omega$.}
	\end{center}
\end{figure}

\begin{figure}[h!]\label{distance}
	\begin{center}
		\includegraphics[scale=.5]{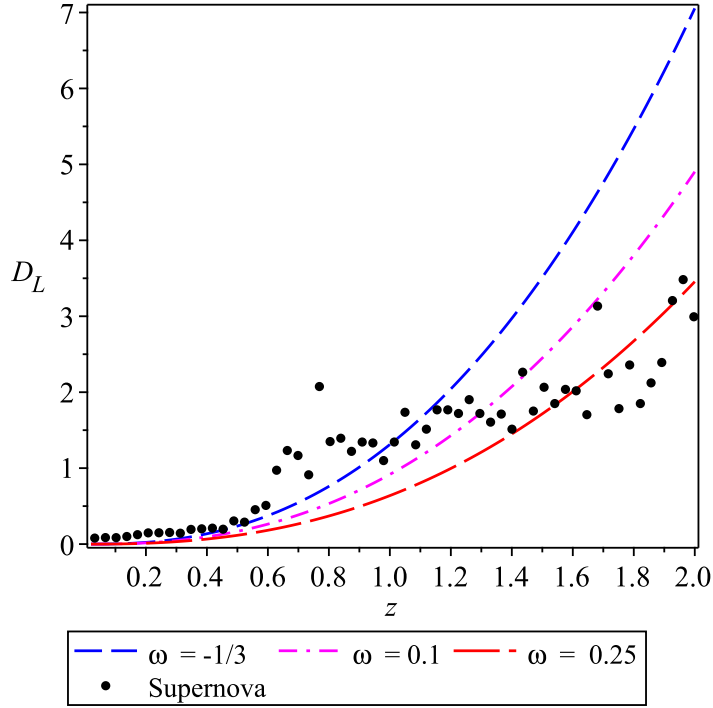}
		\caption{In this figure,  luminosity distance $D_L$ is plotted with respect to redshift $z$ for $\omega$ = -1/3, 0.1 and 0.25. 57 Supernova data are also marked. It is obtained to be positively increasing and consistent with observational data for each $\omega$. }
	\end{center}
\end{figure}

\begin{figure}[h!]\label{ap}
	\begin{center}
		\includegraphics[scale=.5]{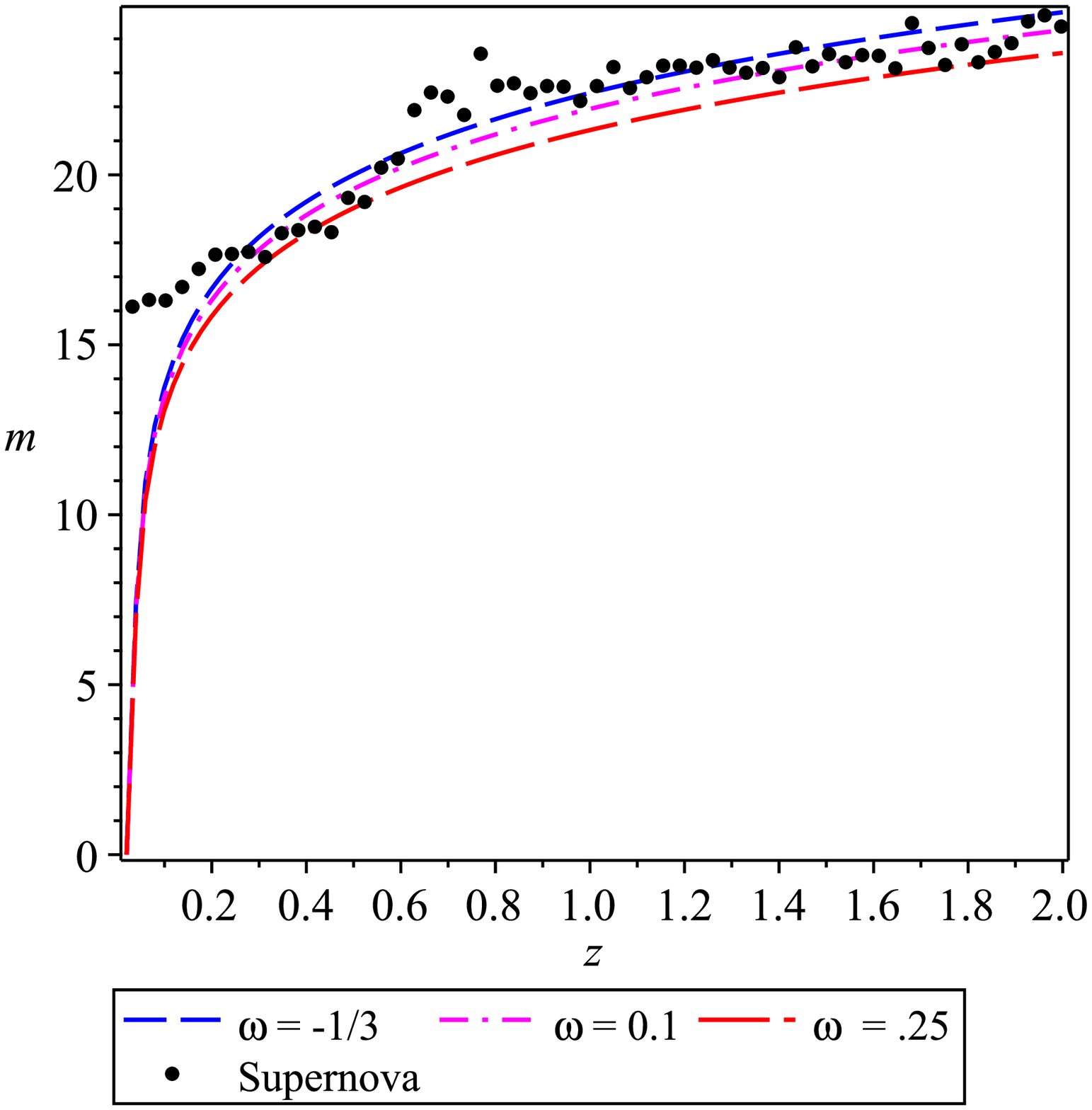}
		\caption{In this figure,   apparent magnitude $m$ is plotted with respect to redshift $z$ for $\omega$ = -1/3, 0.1 and 0.25. 57 Supernova data are also marked. It is obtained to be positively increasing and consistent with observational data for each $\omega$.}
	\end{center}
\end{figure}

\subsection{Energy Conditions}
The energy conditions  to be investigated in the present subsection are: null energy condition (NEC),  weak energy condition (WEC), strong energy condition (SEC) and dominant energy condition (DEC). These are defined in terms of energy density $\rho$ and pressure $p$. If we consider both $\rho$ and $p$ as a function of $z$, then these energy conditions are stated in the following manner:

\noindent
(i) NEC $\Leftrightarrow$ $\rho(z)+p(z)\geq 0$\\

\noindent
(ii) WEC $\Leftrightarrow$ $\rho(z)\geq 0$, $\rho(z)+p(z)\geq 0$\\

\noindent
(iii) SEC $\Leftrightarrow$ $\rho(z)+ p(z)\geq 0$, $\rho(z)+3p(z)\geq 0$\\

\noindent
(iv) DEC $\Leftrightarrow$ $\rho(z)\geq 0$, $\rho(z)-|p(z)|\geq 0$\\

\noindent
\begin{itemize}
	\item  In Eq. \eqref{r}, the energy density is defined in terms of cosmic time. For $\omega =-1$, $\alpha\beta=0$ which implies either $\alpha=0$ or $\beta=0$. If $\alpha=0$, then $\rho(z)$ is constant, which is not true as it is decreasing with the evolution of universe. On the other hand if $\beta=0$, then $\rho(t)$ would be undefined. This restricts that $\omega\in(-5,1/3)-\{-1\}$. Expressing the energy density from \eqref{r} in terms of redshift, we get
\begin{eqnarray}\label{rz}
\rho(z)&=&\frac{8\pi\Big(-1\pm \sqrt{1+h^2(z)}\Big)}{\beta(1-3\omega)(5+\omega)}.
\end{eqnarray}
 For  $\omega=-2, -1/3, 0.1$ and $0.25$, $\rho(z)$ is plotted in Fig. (2). For each of these values, it increases with the increment of $z$ which shows that the energy density is going to decrease as time is increasing.

\noindent
	\item  For our model, we have obtained $\rho(z)+p(z)<0$ for every  $z\geq 0$ with $\omega<-1$ and $\rho(z)+p(z)>0$ for every $z\geq 0$ with $\omega>-1$. This depicts that the universe would be filled with the matter obeying the null energy condition, if $\omega>-1$. Thus, this reduces the range of $\omega$ to (-1,1/3). We
have drawn $\rho(z)+p(z)$ with respect to $z$ in Fig. (3) for $\omega=-2, -1/3, 0.1$ and $0.25$. For $\omega=-2$, it is negative, while for $\omega=-1/3, 0.1$ and $0.25$, it is positive.

\noindent
	\item Clearly, WEC is valid for $z\geq 0$ with $\omega\in (-1,1/3)$.

\noindent
	\item Further, in case of our model SEC term $(1+3\omega)\rho(z)\geq 0$ for all $z\geq 0$ with $\omega<-1$. This implies the non-violation of SEC for  all $z\geq 0$ with $\omega\in (-1,1/3)$. For $\omega=-0.25, 0.1$ and $0.25$, it is plotted in Fig. (4).

\noindent
	\item Furthermore,  DEC term $\rho(z)-|p(z)|=\rho(z)(1-|\omega|)$ is found to be a positive function of $z$ for $\omega$ to (-1,1/3). It is plotted in Fig. (5) for $\omega=-1/3, 0.1$ and $0.25$. Thus, all energy conditions are satisfied for $z\geq 0$ with $\omega \in (-1,1/3)$.

\noindent
\item Thus, our model of universe signifies the presence of matter respecting the energy conditions.
\end{itemize}

\section{Summary and Conclusion}
In the present paper, the framework of FRW model is chosen with the background of $f(Q,T)$ theory of gravity, which has been proposed by Y. Xu et al. in 2019 \cite{Xu} and used to investigate the evolution of FRW model with three specific forms of $f(Q,T)$ gravity models (i) $f(Q,T)=\alpha Q +\beta T$, (ii) $f(Q,T)=\alpha Q^{n+1} +\beta T$ and (iii) $f(Q,T)= -\alpha Q -\beta T^2$. They assumed the universe to be filled with dust matter and obtained the accelerating expansion of the universe. After this study, a natural question arises that what would happen if the dust matter is not filled in the universe or the pressure between the fluid particles is non-zero. To investigate such question, we have considered the non-zero pressure with equation of state $p=\omega \rho$, where $\omega$ is a constant equation of state parameter. Using $f(Q,T)=-\alpha Q-\beta T^2$ model, we have obtained a non-linear differential equation for Hubble parameter in terms of redshift $z$. After re-scaling it in terms of function $h(z)$, we have obtained a non-linear differential equation for $h(z)$ which contains model parameter $k$ and equation of state parameter $\omega$. Using 31 points of Hubble data, we have obtained the constrain on $k$ and obtained its value equal to 1.8 for which $\chi^2$ is minimum. Using this value of $k$, real numerical solution of $h(z)$ is obtained for $\omega\in(-5,1/3)$. It is found to be an increasing function of redshift for each $\omega\in(-5,1/3)$. For lower redshift values, the values of $h(z)$ are closer to the $\Lambda$CDM in comparison of higher redshift values. Further, the energy density $\rho(z)$ is obtained to be undefined for $\omega=-1$ which restricts its range to $(-5,1/3)-\{-1\}$. In order to obtain more constrain, we have found the numerical solution for NEC term $\rho(z)+p(z)$. It is found to be positive for every $z\geq 0$ with $\omega>-1$. Then all SEC and DEC terms are checked and found to be positive for $\omega>-1$. Thus,  we obtained the validity of energy conditions for $z\geq 0$ with  $\omega\in(-1,1/3)$. Then we have calculated the deceleration parameter $q$ which tells about the decelerating or accelerating nature of the universe. For $\omega\in(-1,1/3)$, $-1<q<0$, indicates the accelerating scenario of the universe.  Then we have calculated apparent magnitude and luminosity distance numerically and  used the 57 Supernova data points to examine the consistency between observational and theoretical results. Consequently, the consistency is obtained between the results with respect to each   $\omega\in(-1,1/3)$.

Thus, it is concluded that our model represents the current accelerating and expanding scenario of the universe. It is filled with ordinary matter obeying the  energy conditions and provides the cosmological implications consistent with experimental outcomes. Hence, this work may be fruitful in the further investigation of evolution of our universe.

%

\end{document}